\documentclass[preprint,12pt,1p]{elsarticle}

\usepackage{hyperref}
\hypersetup{
    colorlinks=true,
    linkcolor=blue,
    filecolor=magenta,      
    urlcolor=cyan,
}
\usepackage{amssymb}
 \usepackage{amsthm}
 \usepackage{setspace}
 \usepackage{graphicx}
 \usepackage{listings}
 \usepackage{color}
  \usepackage{bbm}
 \usepackage{amsmath}
\usepackage{amsfonts}
\usepackage{amssymb}
\usepackage{mathrsfs}
\usepackage{graphicx}
\usepackage{setspace}
\usepackage{listings}
\usepackage{color}
\usepackage{subcaption}
\usepackage{mwe}
\usepackage{upquote}%
\usepackage{geometry}
\usepackage{xcolor}

\usepackage{lipsum}
\makeatletter
\def\ps@pprintTitle{%
 \let\@oddhead\@empty
 \let\@evenhead\@empty
 \def\@oddfoot{}%
 \let\@evenfoot\@oddfoot}
\makeatother

\begin{document}

\begin{frontmatter}

\title{Supervised Deep Neural Networks (DNNs)  for Pricing/Calibration of Vanilla/Exotic Options Under Various Different Processes}

\author{Tugce Karatas}
\author{Amir Oskoui}
\author{Ali Hirsa\fnref{myfootnote}}

\fntext[myfootnote]{We would like to thank participants at Shanghai Advanced Institute of Finance,  Shanghai China, Orient Securities Shanghai China, and Cubist Systematic Strategies, New York for their comments and suggestions on this study. Errors are our own responsibility.}




\begin{abstract}
We apply supervised deep neural networks (DNNs) for pricing and calibration of both vanilla and exotic options under both diffusion and pure jump processes with and without stochastic volatility. We train our neural network models under different number of layers, neurons per layer, and various different activation functions in order to find which combinations work better empirically. For training, we consider various different loss functions and optimization routines. We demonstrate that deep neural networks exponentially expedite option pricing compared to commonly used option pricing methods which consequently make calibration and parameter estimation super fast.
\end{abstract}


\end{frontmatter}


\section{Introduction}

A major component of model calibration and parameter estimation is derivatives pricing. Aside from few cases there is no closed-form solution for derivatives pricing. In computational finance, it is critical to find a fast and accurate approximation  for cases that there is no analytical solution. Some of these approaches could be very time-consuming and to possess a model that is accurate and fast in pricing is central. Movement in time can make the calculated values obsolete. Hence, the time it takes to re-compute these values is pivotal in derivatives business. Sometimes it is possible to exploit redundancies in the calculations: for example, pricing via fast Fourier techniques (FFT) \cite{FFT}, which results in a vector of prices on the same underlier for different strikes. To do so, it takes advantage of redundant matrix operations and cuts down on computational cost. In the setting of neural networks, once training is done, the mapping from input parameters: $\bf{X}$ to an output price $\bf{y}$ has been optimized to give the smallest loss, through many iterations of updates on weights and biases in combination with permutations of non-linear activation functions. Hence, with its trained architecture the calculation of prices using neural networks both with and without feedback connections which mimic the idea of context or memory in the brain (i.e. recurrent and feedforward neural networks respectively), becomes elementary and at the same time super-fast.\\

In this paper, we revisit the original work of Das and Culkin \href{https://srdas.github.io/Papers/BlackScholesNN.pdf}{(2017)} that used deep neural  networks  for  option  pricing  and  explore  various  architectures  and   extend the work to more advanced models and different contracts. We show how Feedforward Neural Networks can be used in the context of derivative pricing to achieve speed-ups of many orders of magnitude. As are the laws of conservation, the speed-up comes with a cost of accuracy. However, we show the loss in accuracy is well within a worst case estimate of bid-ask spreads, making the method practical.  We demonstrate how to create labels to train the model for vanilla and exotic options (with weak and strong path dependence) under diffusion and pure jump processes with and without stochastic volatility \cite{Heston, CGMY, American}. To optimize speed and accuracy of the Neural Network's learning we use many perturbations of hyper-parameters (i.e. number of layers, neurons per layer, and activation functions). In particular we show how European, Barrier, and American option prices can be trained and priced out-of-sample swiftly and accurately under geometric Brownian motion (GBM) \cite{Black} and variance gamma (VG) \cite{VG}, both with and without stochastic arrivals \cite{Heston, CGMY}. We show that this method trumps traditional techniques for option pricing which results in super-fast calibration and parameter estimation.

\subsection{Introduction to Models}
To model market behavior we used four different processes: (a) geometric Brownian motion (GBM) \cite{Black}, (b) geometric Brownian motion with stochastic arrival (GBMSA) a.k.a Heston stochastic volatility model \cite{Heston}, (c) variance gamma (VG) model \cite{VG}, and (4) variance gamma with stochastic arrival (VGSA)\cite{CGMY}.\\

The first two are pure diffusion processes with constant and stochastic volatility respectively. The last two are pure jump processes, mirroring the diffusion cases, VG has constant volatility whereas  VGSA has stochastic arrivals to allow for volatility clustering \cite{CGMY}. GBM as in the Black-Merton-Scholes model is the most well known for derivatives pricing \cite{Black}. However, its assumption of a constant volatility does not reflect market behavior, leading to an implied volatility surface that facilitates model inputs to match market prices. The variance gamma process is a pure-jump process with time-changed Brownian motion to account for high activity \cite{VG}. It is shown that with addition of two extra parameters skewness and kurtosis we can closely fit the smile across strikes for a fixed maturity \cite{VG}. Both GBMSA and VGSA are stochastic volatility models for diffusion and jump processes respectively, with an effort to model non-constant volatility, as is observed in the market, enabling "us to calibrate to option price surfaces across both strike and maturity simultaneously" \cite{Hirsa}.

\section{Brief Introduction to Neural Networks}
\subsection{Feedforward Neural Networks}
Here, we will briefly introduce feedforward neural networks but for a more detailed discussion of DNNs we refer the reader to chapter 6 of: \href{http://www.deeplearningbook.org/contents/mlp.html
}{Deep Learning by Goodfellow, Bengio and Courville }. Feedforward neural networks, also called deep feedforward networks, are considered "quintessential" to deep learning with the goal of approximating a function: {\bf${f^*}$} \cite{Goodfellow}. Where the network defines a mapping $\boldsymbol{y} = f(\boldsymbol{x};\boldsymbol{\theta})$ and learns the appropriate parameter set $\boldsymbol{\theta}$ that leads to an optimal function approximation. The network is basically a composition of many different functions in a chain. The more of these functions that are applied on your input data $\bf{X}$, the more $\bf{layers}$ in your network. The length of functions compositely applied is in a sense the $\bf{depth}$ of the model. These networks must be trained in order to mimic a function or model.\\

Training data is provided accompanying the input $x$ so that the learning algorithm can decide how to use its layers to produce values as close as possible to $\hat{y}$. During training we have an optimizer that drives the output of the neural network $y$ = $f(x)$ to match $\hat{y}$ = $f^*(x)$. Hence, the neural network learns to use its layers to minimize an $\bf{objective}$ $\bf{function}$, in our case a Mean-Squared-Error function:

\begin{center}
\begin{equation*}
MSE  = \displaystyle\frac{1}{N}\sum_{i=1}^{N}(y-\hat{y})^2
\end{equation*}
\end{center}
The training data only indicates what values the output layer must approximate for each input $x$; however, it does not directly indicate how to use the layers to optimally approximate $f^*$. When initially looking at training data it is not clear what the appropriate output for each layer should be, which is why they are referred to as $\bf{hidden}$ $\bf{layers}$ in a "neural" network.\\

\begin{figure}[!h]
    \centering
    \includegraphics[width=\textwidth]{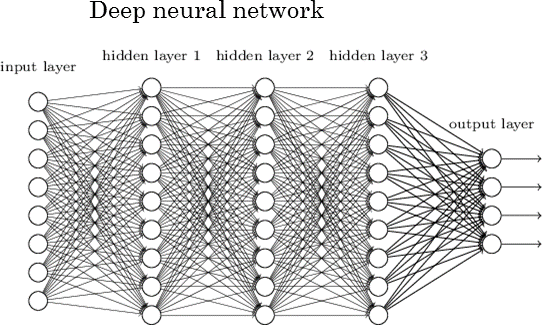}
    \caption{A graphical visualization of a Deep Neural Network from \href{https://www.rsipvision.com/exploring-deep-learning/
}{here}}
    \label{fig:DNNGraph}
\end{figure}

The term neural comes from a loose inspiration of neuroscience and the dimensionality of these hidden layers. As, the width of each layer is based on the number of elements within it that are mapping a vector to a scalar via a non-linear activation function. In that sense each of these elements is analogous to a $\bf{neuron}$, because they are taking inputs from many different things and applying their own activation function \cite{Goodfellow}. A visualization of both depth and width in a feed-forward neural network can be seen in Figure \ref{fig:DNNGraph}.\\

\subsection{Convolutional Neural Networks}
Another way of specializing neural networks to replicate complex functions in high dimensions is convolutional neural networks, also referred to as CNNs \cite{LeCun}. They often work well with data that "has a clear grid-structured topology" like time-series data or image data which are 1-D and 2-D grids respectively \cite{Goodfellow}. Their huge success in practical applications for high dimensional functions leads us to believe it could be tremendously useful and intuitive to use in the context of Asian Options as they have a grid-like topology \cite{Asian}.

\subsection{"Wide" vs "Deep"}

\begin{figure}[!ht]
    \centering
    \includegraphics[width=\textwidth]{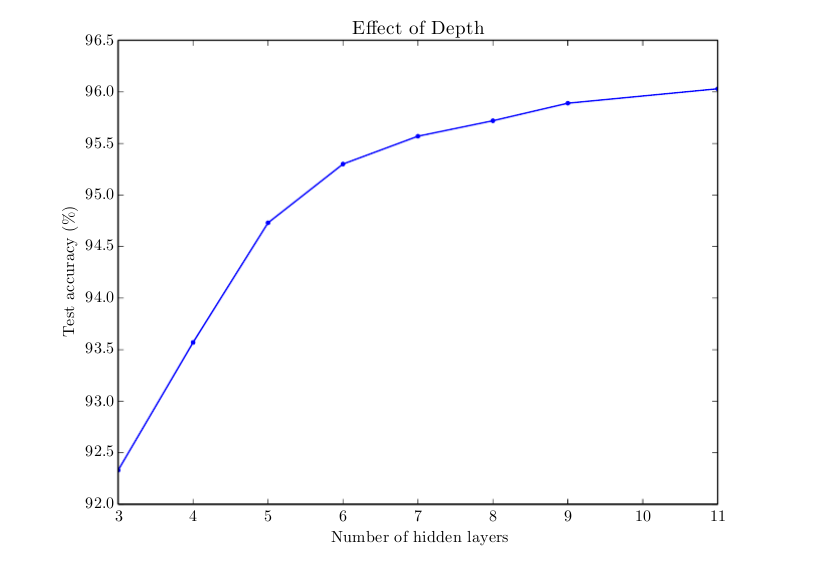}
    \caption{Empirical results showing that deeper networks generalize better when used to transcribe multi-digit numbers from photographs of addresses. Data from Goodfellow \textit{et al.} (2014d). The test set accuracy consistently increases with increasing depth...\cite{Black}}
    \label{fig:EffectOfDepth}
\end{figure}

In Figure \ref{fig:EffectOfDepth} there is a visualization of the "Effect of Depth" showing how an increased number of layers in a neural network (i.e. deepness) results in higher accuracy of prediction shown by Goodfellow \textit{et al.} (2014d). This seems like a compelling point that favors deepness of the neural network; however, Zogoruyko and Komadakis \href{https://arxiv.org/abs/1605.07146}{(2017)} noted that by doubling the number of layers accuracy was only improved by a fraction of a percent, leading to a ''problem of diminishing feature reuse, which makes these networks very slow to train." They demonstrated how a ''wide" neural network (meaning the number of hidden units is at-least twice the input dimension) of only 16-layers could outperform ''deep" neural networks including ''thousand-layer-deep networks, achieving new state-of-the-art results on CIFAR, SVHN, COCO and significant improvements on ImageNet." In the context of derivative pricing we have also explored these two compelling arguments and the results can be seen in Figure \ref{Fig:Data6}. This result clearly highlighted the diminishing accuracy of the networks with an increase in depth, resulting in us to explore an optimal width for our hidden layers as shown in Figure \ref{Fig:Data5} which will be discussed further.

\subsection{Dropout}
After exhausting the width and depth dimensions of our neural network, another idea worth exercising is dropout. To prevent over-fitting of the model we can force our neural network to learn the more robust features in conjunction with many different random subsets of the other neurons. This is called dropout because nodes of the network are dropped with a  probability 1-$p$ at each training stage.

\subsection{Choice of Optimizer and Global vs Local Convergence}
Once this structure has been integrated in the neural network, optimization can be  done using a choice of either gradient descent or more sophisticated routines like RMSProp, Adam or stochastic gradient descent (SGD) with momentum. Hence, as for deciding the choice of optimizer in this architecture we introduce some insights, both local and global, studied by Soltanolkotabi et al. (2018) for reaching optima on wide and over-parameterized neural networks\cite{Soltanolkotabi}. For a 2-layered neural network with Leaky ReLU activation function, Soudry and Carmon (2016) have shown that a global minimum can be obtained using gradient descent on a modified loss function; however, this doesn't evince that the global minimum of the original loss function is reached \cite{Soudry}. Li and Yuan (2017) have introduced identity mapping, by which SGD always converges to the global minimum for a 2-layered neural network with ReLU activation function under the standard weight initialization scheme \cite{li2017convergence}. Under similar realistic assumptions, Kawaguchi's studies showed that all local minima are global minima using nonlinear activation functions \cite{Kawaguchi}. Additionally, under several assumptions Choromanska et al. (2015) have simplified the loss function to a polynomial with i.i.d. Gaussian coefficients, showing that using stochastic gradient descent all critical points found "are local minima of high quality" meaning the test error value is comparable to that at the global minima \cite{Choromanska}. They concluded that global minima are more difficult to reach with larger networks, but in practice it's "irrelevant as global minimum often leads to overfitting" \cite{Choromanska}.

To further generalize previous studies, Kawaguchi et al. (2018) show that without the over-parameterization or any simplification assumption, local minima of the DNN are theoretically "no worse than the globally optimal values of corresponding classical machine learning models" \cite{Kawaguchi2}. This theoretical observation was supported empirically using stochastic gradient descent with momentum as their optimizer during training for ReLU, Leaky ReLU, and absolute value activation functions\cite{Kawaguchi2}. Around the same time, Allen-Zhu et al. (2018) noted that under the non-degenerate input and over-parameterization assumption, global optima can be reached using simple algorithms such as SGD during training of the DNN, but this result cannot be extended to testing \cite{Allen-Zhu}. Furthermore, Kingma and Ba (2014) have introduced Adam (Adaptive Moment Estimation) optimization routine and they have shown that Adam converges faster compared to SGD and RMSProp when a neural network with ReLU activation function is trained \cite{kingma2014adam}. These theoretical and empirical findings lead us to use more sophisticated optimization routines rather than gradient descent.

For our problem, we first used stochastic gradient descent (SGD) as the optimization routine which was painfully slow in training the model as expected. We then utilized RMSprop and Adam optimization routines since they are both commonly used by practitioners. Adam was of particular interest because it combines advantages from both AdaGrad and RMSProp, which are both extensions to stochastic gradient descent. Based on our empirical observations, both RMSprop and Adam converges much faster than SGD. Furthermore, Adam seems to work better than RMSprop, in parallel with what was expected from its bias-correction term and the insights mentioned above for an over-paramaterized DNN with ReLU, Leaky ReLU, and absolute value activation functions\cite{kingma2014adam, Ruder}. Thus, for the remainder of this paper, Adam is used as our optimization routine for training the network. 





\section{Labels and Parameter Selection}
In order to implement supervised deep neural networks, many labels are needed for training. These labels are generated from existing\footnote{By existing models, we mean analytical or computational  models used for derivatives pricing e.g. transform techniques, numerical solutions of PDEs or PIDEs, Mote-Carlo simulation and the like} models. Table \ref{table:metric} shows some classical techniques for pricing options and it shows under a given model, which types of options that can be used for pricing. Monte-Carlo simulation is commonly considered the slowest amongst these methods and when it comes to trees, they have an order of complexity that is quadratic with each time step. Now, even though discretization may lead Monte Carlo simulation in terms of speeds and they both have a very wide domain of option types and models where they can be used, they are both still considered computationally expensive in practice. Leaving us with the transform techniques like the FFT method, which may be named ''Fast", but still takes a reasonable amount of  time  to  run and these transforms are not applicable for path-dependent options. This results in the crux of our problem, in other words the most expensive and computationally time-consuming step when using supervised deep neural networks, is generating labels from these existing methodologies. Once labels are generated, training feedforward neural networks becomes almost trivial.  
\begin{table}[htbp]
\begin{center}
\scalebox{0.75}{
\begin{tabular}{|c||c|c|c||c|c|c||c|c|c|||c|c|c|} \hline
 & \multicolumn{12}{|c|}{Pricing Model/Method} \\ \cline{2-13}
 & \multicolumn{3}{|c||}{{\small Transform Techniques}}& \multicolumn{3}{|c||}{{\small PDEs/PIDEs}}&\multicolumn{3}{|c|||}{{\small MC simulation}}&\multicolumn{3}{|c|}{{\small DNN}} \\\cline{2-13}
  \raisebox{2.0ex}[0pt]{model} & {\small vanilla} & {\small weak} & {\small exotic}& {\small vanilla} & {\small weak} & {\small exotic}& {\small vanilla} & {\small weak} & {\small exotic}& {\small vanilla} & {\small weak} & {\small exotic}  \\ \hline
 GBM & \checkmark & \checkmark & \color{blue}$\times$ & \checkmark & \checkmark & \checkmark & \checkmark & \checkmark & \checkmark& \color{red}\checkmark &  \color{red}\checkmark & \color{red}\checkmark \\
 VG & \checkmark & \checkmark & \color{blue}$\times$ & \checkmark & \checkmark & \checkmark & \checkmark & \checkmark & \checkmark & \color{red}\checkmark & \color{red}\checkmark & \color{red}\checkmark\\ \hline
 GBMSA   & \checkmark & \checkmark & \color{blue} $\times$ & \checkmark & \checkmark & \checkmark & \checkmark & \checkmark & \checkmark & \color{red}\checkmark & \color{red}\checkmark & \color{red} ?\\
 VGSA  & \checkmark & \checkmark & \color{blue} $\times$ & \checkmark & \checkmark & \checkmark & \checkmark & \checkmark & \checkmark & \color{red}\checkmark & \color{red}\checkmark &  \color{red} ? \\
\hline
\end{tabular}}
\end{center}
\caption{Pricing schemes for various different models/processes}
\label{table:metric}
\end{table}

After deciding on which method and model to use for each option, the next step is to decide on market and model parameter ranges. Before going through model parameters for each option, we first start by deciding on the range of market parameters, which will be the same throughout all models and options. As we stated before, supervised deep neural networks necessitates creating a large number of labels from existing models. There are a general set of parameters that are common amongst all of these models: initial stock price ($S_0$), strike price (K), maturity (T), interest rate (r), dividend yield (q). Using the fact that option prices are linear homogeneous in $S_0$ and K as seen in Black-Merton-Scholes option pricing formula, instead of considering both $S_0$ and K as two separate parameters, we will only consider the ratio $S_0/K$ that is the option moneyness. Table \ref{table:marketparameters} shows the product and market parameters we have considered for each model. For training, we generated labels within 20\% moneyness. The range for the maturity of the options is from 1 day to 3 years. 

\begin{table}[htbp]
\vskip\baselineskip 
\begin{center}
\begin{tabular}{l l}\hline
\textbf{Parameter}& \textbf{Range}\\\hline
Moneyness($S_0/K$) & 0.8 $\rightarrow$ 1.2  \\
Maturity(T) & 1 day $\rightarrow$ 3 years  \\
Dividend rate(q) & 0\% $\rightarrow$ 3\%  \\
Risk free rate(r) & 1\% $\rightarrow$ 3\%  \\ \hline
\end{tabular}
\end{center}
\caption{Product and market parameters}
\label{table:marketparameters}
\end{table}
\subsection{European Options}
In the case of Vanilla Options, there are many available methods for pricing including Fast Fourier Transform (FFT), Trees, and Monte-Carlo Simulation.  Leading us to build a Deep Neural Network (DNN) trained on closed-form generated call and put prices from the Black-Merton-Scholes formula for the GBM model. For the VG, VGSA, and GBMSA models, the Deep Neural Network is trained using labels obtained via FFT considering we have the characteristic function of the logarithm of the stock price process under these models.

The process begins with creating an input parameter matrix {\bf X}, which consists of different combinations of market and model parameters. Here, model parameters are volatility ($\sigma$) for the GBM model; volatility ($\sigma$), skewness ($\theta$) and kurtosis ($\nu $) for the VG model; rate of mean reversion ($\kappa$), correlation between stock and variance ($\rho$), long term variance ($\theta$), volatility of variance ($\sigma$) and the initial variance ($v_0$) for the GBMSA model; and volatility ($\sigma$), skewness ($\theta$), kurtosis ($\nu$), rate of mean reversion ($\kappa$), the long-term rate of time change ($\eta$), and the volatility of the time change ($\lambda$) for the VGSA model. Table \ref{table:modelparameters} shows the ranges of model parameters we have considered for European options. 

After determining the ranges for the parameters, the parameter combinations are sampled via two different approaches. The first one is Non-Quasi approach, where parameter combinations are sampled at random, uniformly over the given ranges of market and model parameters. The second approach is the Quasi approach. For this approach, we benefit from Halton sequences, which are deterministic and of low-discrepancy. Here, we considered a 5-dimensional Halton sequence for the GBM model, 7-dimensional for the VG model, 9-dimensional for GBMSA model, and 10-dimensional for the VGSA model. For this approach, Halton sequences are generated in the R platform as a first step. Then, the value of each parameter is obtained by multiplying the numbers, which are obtained from Halton sequences, with the range of the parameter and summing it with the lower bound of the corresponding parameter.

After the parameter ranges and sampling methods are chosen, the next step is to decide on the training size. Here, we considered 4 different training sizes, which are 40,000, 80,000, 160,000, and 240,000, in order to observe the effect of the training size on the performance. Each data set is generated separately. As expected, the discrepancy between data points decreases as training size increases. For this reason, it is expected that the performance of the model increases as training size increases. 

\begin{table}[htbp]
\vskip\baselineskip 
\begin{center}
\scalebox{0.90}{
\begin{tabular}{c l c l c l c l }\hline
\textbf{GBM}& \textbf{Range} & \textbf{VG}& \textbf{Range} & \textbf{GBMSA}& \textbf{Range} & \textbf{VGSA}& \textbf{Range}\\\hline
$\sigma$ & 0.05 $\rightarrow$ 0.50 & $\sigma$ & 0.05 $\rightarrow$ 0.50 & $\sigma$ & 0.05 $\rightarrow$ 0.50 & $\sigma$ & 0.05 $\rightarrow$ 0.50 \\
& & $\theta$ & -0.90 $\rightarrow$ -0.05 & $\kappa$ & 0.20 $\rightarrow$ 2.00 & $\theta$ & -0.90$\rightarrow$-0.05 \\
& & $\nu$ & 0.05 $\rightarrow$ 1.00 & $\rho$ & -0.90 $\rightarrow$ -0.10 & $\nu$ & 0.05 $\rightarrow$ 1.00\\
&&&&$\theta$ & 0.01 $\rightarrow$ 0.20 & $\kappa$ & 0.20 $\rightarrow$ 3.00  \\
&&&&$v_0$ & 0.01 $\rightarrow$ 0.20 & $\eta$ & 0.01 $\rightarrow$ 0.20  \\
&&&& &  & $\lambda$ & 0.01 $\rightarrow$ 0.20  \\\hline
\end{tabular}}
\end{center}
\caption{Model parameters}
\label{table:modelparameters}
\end{table}

Once the input parameter matrix ${\bf X =[S_0/K, T, r, q, \boldsymbol{\sigma}]}$ is produced for the GBM model, the output vector ${\boldsymbol{y}}= {\bf EC/K}$ is obtained by using the closed form Black-Merton-Scholes formula, where {\bf EC} is the price of European call option. For VG, GBMSA, and VGSA models, the output vector ${\boldsymbol{y}}= {\bf EC/K}$ is obtained by using the FFT algorithm for corresponding input parameter matrices {\bf X}. Therefore, $({\bf X},{\boldsymbol{y}})$ will form the training set of the deep neural network.

\subsection{Barrier Options}
To introduce some path dependence, we transition implementation to Barrier Options; for which, closed-form solutions are available in the GBM setting. However, prices are usually computed using Monte-Carlo simulation for beyond GBM\footnote{Depends on the payoff, we can apply transform techniques like the Cosine-Fourier (COS) method \cite{fang2009pricing} for pricing some barrier options, however for this study we tried Monte-Carlo methods for pricing.}. In this section, we will focus on pricing an Up-and-Out Put Options (UOP), where the Variance Gamma and Geometric Brownian Motion models are used to describe the market both with and without stochastic arrivals. The ranges of all product and market parameters are the same with that of European options as in Table \ref{table:marketparameters} except the range of {\bf H/K}, where {\bf H} is the barrier level. Since we used the normalized values for stock prices and option prices, we consider the normalized values of barrier levels as well. For {\bf H/K}, we consider the range between ${\bf S_0/K}$ and 1.2, where 1.2 is the upper bound for moneyness considered before. The reason behind the lower bound of this range is that the barrier level must be greater than the initial stock price for up-and-out barrier options. Furthermore, the ranges of the model parameters come from Table \ref{table:modelparameters} as in European options. Again, the parameter combinations are sampled at random uniformly over the given ranges of market and model parameters.

Once the input parameter matrix ${\bf X =[S_0/K,H/K, T, r, q, \boldsymbol{\sigma}]}$ is produced for the GBM model, the output vector $\bf \boldsymbol{y} = UOP/K$ is obtained by using the closed form solution. Here, {\bf UOP} is the price of up-and-out put options. For the GBMSA and VGSA models the output vector $\bf \boldsymbol{y} = UOP/K$ is obtained by using Monte Carlo procedure for their corresponding input parameter matrices {\bf X}. This procedure is based on 10,000 price paths of the underlying for the GBMSA model and 8,000 price paths of the underlying for the VGSA model where each simulated path has 100 time-steps. For details on how to simulate GBMSA or VGSA processes, see \cite{Hirsa} page 239.

\subsection{American Options}
Now to really get to see where neural networks can transform the norm for pricing times, we introduce the strong path dependence of American Options to this architecture. American Options are usually priced using either Monte-Carlo simulation or tree methods. Since both of these methods are relatively time consuming, we construct a fully-connected feedforward neural network trained on the prices of American Options generated via the Ju-Zhong (JZ) approximation in the GBM model. With input parameter matrix: ${\bf X =[S_0/K, T, r, q, \boldsymbol{\sigma}]}$, the output vector ${\bf \boldsymbol{y} = AJZ/K}$ is obtained by using the Ju-Zhong Approximation. Here, {\bf JZ} is the price of American Put Options under the Ju-Zhong Approximation. Therefore, $({\bf X},{\boldsymbol{y}})$ will form the training set of the deep neural network. While generating input parameter matrices for American options, we used the Non-Quasi approach, which is described in Section 3.1. Therefore, market parameter combinations are generated at random uniformly over the ranges provided in Table \ref{table:marketparameters}, and the model parameter $\sigma$ is generated from a uniform random distribution over the range between 0.05 and 0.50.


\section{Training}
Once training sets $({\bf X},{\boldsymbol{y}})$ are obtained for each model and option type, they are fed to the deep neural network algorithm. We train each model under a different number of layers, a different number of neurons per layer and various different activation function combinations in order to observe which combinations work better empirically under given training sets. 

\subsection{Structure/Architecture}

Non-linearity in the model is introduced through activation functions. We use different combinations of activation functions. In this experiment, we have considered exponential linear unit (elu), rectified linear unit (relu), leaky rectified linear unit (leaky relu), and sigmoid activation functions. For both sigmoid and tanh functions if the input is either very large or small, the slope of the activation becomes very small resulting in a slow down for gradient descent making either of these functions not well suited for hidden layers \cite{Glorot10understandingthe}. However, to not rule anything out we have still stress tested permutations of these activation functions and visualized the results. There is a common preference towards rectified linear unit (relu) for hidden layers as its derivative is 1 for all positive inputs and 0 for negative ones, and to avoid a zero derivative in many cases leaky relu can be substituted \cite{Glorot2}.

\begin{figure}[!htb]
   \begin{minipage}{0.5\textwidth}
     \centering
     \includegraphics[width=.9\linewidth]{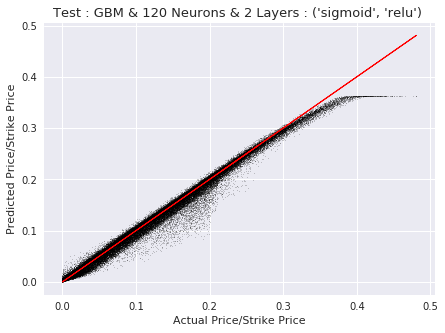}
     \caption{Sample-1}\label{Fig:Data1}
   \end{minipage}\hfill
   \begin{minipage}{0.5\textwidth}
     \centering
     \includegraphics[width=.9\linewidth]{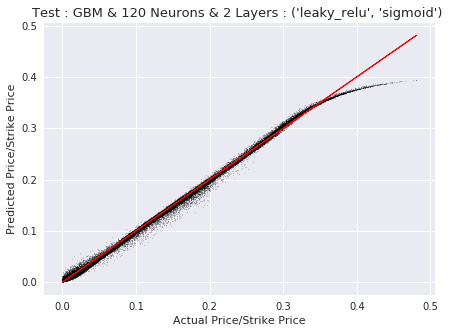}
     \caption{Sample-2}\label{Fig:Data2}
   \end{minipage}
\end{figure}

\begin{figure}[!htb]
   \begin{minipage}{0.5\textwidth}
     \centering
     \includegraphics[width=.9\linewidth]{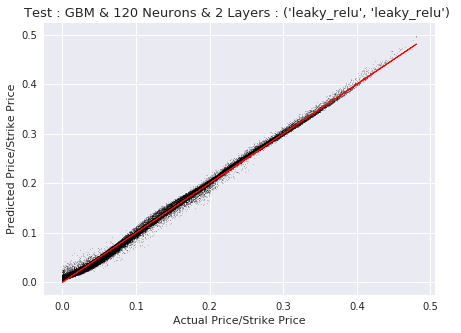}
     \caption{Sample-3}\label{Fig:Data3}
   \end{minipage}\hfill
   \begin{minipage}{0.5\textwidth}
     \centering
     \includegraphics[width=.9\linewidth]{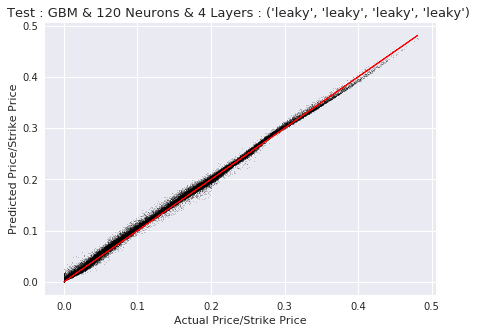}
     \caption{Sample-4}\label{Fig:Data4}
   \end{minipage}
\end{figure}

As we expected, Figures \ref{Fig:Data1} and \ref{Fig:Data2} illustrate that a sigmoid activation function is not suitable for our structure and going forward we will omit using it in our model and continue with the combinations of elu, relu, and leaky relu activation functions. After testing all possible combinations of these three activation functions for 2, 3, and 4 number of layers, we realized that the trained models with initial activation function of leaky relu perform better compared to other trained models. Figures \ref{Fig:Data3} and \ref{Fig:Data4} are good examples for this argument. This argument is not only valid for GBM model, but it is true for other processes as well. Based on these empirical observations, we will set the initial activation function as leaky relu, and test different combinations of activation functions for other hidden layers. We will then select the trained model with the smallest MSE value or the largest $R^2$ value as our pricing neural net.

\subsection{Diagnostics}
\begin{figure}[!htb]
\begin{center}
\includegraphics[scale = 0.6]{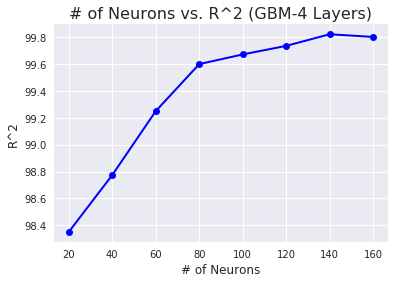}
\caption{Number of Neurons vs. $R^2$}\label{Fig:Data5}
\end{center}
\end{figure}

After testing different combinations of activation functions, the next step is to test the effect of a different number of layers as well as a different number of neurons per layer. As the first experiment, we keep the number of layers and the set of activation functions fixed, and vary the number of neurons per layer. Figure \ref{Fig:Data5} shows how the $R^2$ value changes as the number of neurons per layer increases. Here, we train Deep Neural Network with European call option prices under the GBM model. There are 4 hidden layers, and the activation function is leaky relu for each hidden layer. 
We can see in Figure \ref{Fig:Data5} that as the number of neurons per layer increases, the neural net model performs better, but there are diminishing returns. After 120 neurons per layer, there is a plateauing result, making it pointless to increase the number of neurons per layer further. Hence, we used 120 neurons per layer as default. 

\begin{figure}[!htb]
\begin{center}
\includegraphics[scale = 0.6]{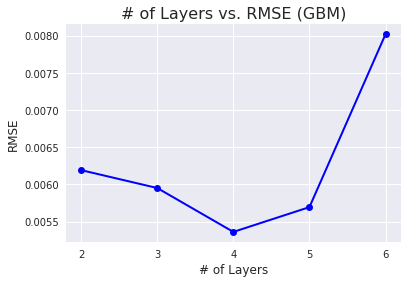}
\caption{Number of Layers vs. RMSE}\label{Fig:Data6}
\end{center}
\end{figure}

In Figure \ref{Fig:Data6}, we examine the model performance as the number of layers is increased. We show how RMSE (root mean square error) changes when we increase the number of layers for European call options under the GBM model. This finding is pretty consistent for other models. 
It can be seen in Figure \ref{Fig:Data6} that as the number of layers increases, the model performance gets better. But this improvement is limited to 4 layers. After that, as soon as we use 5 layers or higher, RMSE value increases. It can be interpreted in a way that as we increase the number of layers, after some point there will be an over-parametrization problem. Based on this observation we cap the number of layers in our neural network to four.
\section{Validation}
Once we obtain the optimal neural network architecture for each model and each option type, we need to check the architecture's validity. In order to do so, we check our trained model for three cases: (1) interpolation, (2) deep-out-of-the-money, and (3) longer maturity. In the first case, we just validate it for interpolated points on our testing set. In the second case, we keep everything else fixed, and test how the trained models behave when $S_0/K$ is between 0.6 and 0.8. In the third case, we keep everything else fixed, and test how the trained models behave when the maturity is between 3 years and 5 years. 
\subsection{European Options}
European call option prices are generated under 4 different models; hence, we obtain 4 different neural network architectures for European call options. Their performances are measured in terms of MSE (mean square error) and $R^2$ values. Table \ref{table:EC-1} provides the summary of the empirical results for the trained models under GBM and VG, and Table \ref{table:EC-2} provides the summary of the empirical results for the trained models under the GBMSA and VGSA models. For all these tables, the in-sample test size and the test sizes for deep-out-of-the-money and longer maturity cases are taken as 60,000. 
\begin{table}[!htb]
\begin{center}
\scalebox{0.8}{
\begin{tabular}{|c|c|c|c|c||c|c|c|c|}\hline
& \multicolumn{4}{|c||}{{\small \textbf{GBM}}}&  \multicolumn{4}{|c|}{{\small \textbf{VG}}}\\\cline{2-9}
{\small \textbf{Training set size}}&40000&80000&160000&240000&40000&80000&160000&240000 \\ \hline
\multicolumn{1}{|c|}{{\small \textbf{In-Sample}}} & \multicolumn{4}{|c||}{}&  \multicolumn{4}{|c|}{}\\\hline
MSE & 0.000060 & 0.000034 & 0.000029& 0.000018 & 0.000678 & 0.000059 & 0.000046 & 0.000028\\\hline
$R^2$ &  99.32\% & 99.62\% & 99.67\%& 99.79\% & 94.18\% & 99.50 \% & 99.61 \% & 99.76 \% \\\hline
\multicolumn{1}{|c|}{{\small \textbf{Deep-Out-Of-The-Money}}} & \multicolumn{4}{|c||}{}&  \multicolumn{4}{|c|}{}\\\hline
MSE & 0.000065 & 0.000054 & 0.000022& 0.000014 & 0.000376 & 0.000071 & 0.000050 & 0.000042\\\hline
$R^2$ &  95.37\% & 96.17\% & 98.47\%& 98.70\% & 83.78 \% & 96.92 \% & 97.83 \% & 98.17 \% \\\hline
\multicolumn{1}{|c|}{{\small \textbf{Longer Maturity}}} & \multicolumn{4}{|c||}{}&  \multicolumn{4}{|c|}{}\\\hline
MSE & 0.000348 & 0.000184 & 0.000191& 0.000129 & 0.004338 & 0.000643 & 0.000524 & 0.000378\\\hline
$R^2$ &  97.34\% & 98.59\% & 98.54\%& 99.01\% & 69.03 \% & 95.41\% & 96.26 \% & 97.30\%\\\hline
\end{tabular}}
\end{center}
\caption{European Option Pricing under GBM and VG}
\label{table:EC-1}
\end{table}

According to the results in Tables \ref{table:EC-1} and \ref{table:EC-2}, it can be observed that as the training size increases, the performance of the trained models increases but there are diminishing returns. The second result is that the performance of the trained neural net under the GBM model is better than all other processes. This is probably because as the number of parameters increases, the process becomes more complicated to learn and probably more labels are needed to train them.
\begin{table}[!htb]
\begin{center}
\scalebox{0.80}{
\begin{tabular}{|c|c|c|c|c||c|c|c|c|}\hline
& \multicolumn{4}{|c||}{{\small \textbf{GBMSA}}}&  \multicolumn{4}{|c|}{{\small \textbf{VGSA}}}\\\cline{2-9}
{\small \textbf{Training set size}}&40000&80000&160000&240000&40000&80000&160000&240000 \\ \hline
\multicolumn{1}{|c|}{{\small \textbf{In-Sample}}} & \multicolumn{4}{|c||}{}&  \multicolumn{4}{|c|}{}\\\hline
MSE & 0.000122 & 0.000059 & 0.000032& 0.000025 & 0.000240 & 0.000230 & 0.000244 & 0.000092\\\hline
$R^2$ &  98.48\% & 99.22\% & 99.60\%& 99.69\% & 98.23\% & 98.31 \% & 98.21 \% & 99.32 \% \\\hline
\multicolumn{1}{|c|}{{\small \textbf{Deep-Out-Of-The-Money}}} & \multicolumn{4}{|c||}{}&  \multicolumn{4}{|c|}{}\\\hline
MSE & 0.000066 & 0.000032 & 0.000021& 0.000016 & 0.000226 & 0.000324 & 0.000271 & 0.000150\\\hline
$R^2$ &  93.48\% & 96.88\% & 97.97\%& 98.46\% & 94.00 \% & 91.38 \% & 92.79 \% & 96.00 \% \\\hline
\multicolumn{1}{|c|}{{\small \textbf{Longer Maturity}}} & \multicolumn{4}{|c||}{}&  \multicolumn{4}{|c|}{}\\\hline
MSE & 0.000362 & 0.000237 & 0.000158& 0.000103 & 0.000647 & 0.000724 & 0.000755 & 0.000377\\\hline
$R^2$ &  96.36\% & 97.61\% & 98.41\%& 98.97\% & 95.92\% & 95.43 \% & 95.23 \% & 97.62\%\\\hline
\end{tabular}}
\end{center}
\caption{European Option Pricing under GBMSA and VGSA}
\label{table:EC-2}
\end{table}

\begin{figure}[!htb]
   \begin{minipage}{0.5\textwidth}
     \centering
     \includegraphics[width=0.9\linewidth]{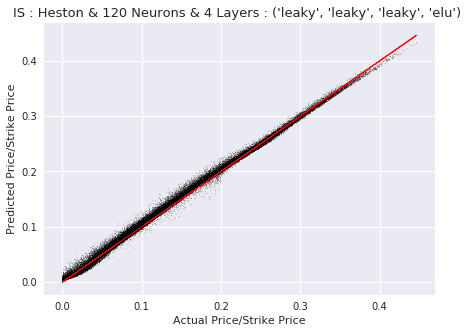}
   \end{minipage}\hfill
      \begin{minipage}{0.5\textwidth}
     \centering
     \includegraphics[width=0.9\linewidth]{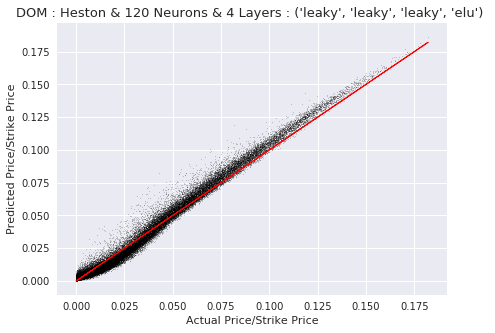}
   \end{minipage}\hfill
   \begin{center}
   \begin{minipage}{0.5\textwidth}
   \centering
     \includegraphics[width=0.9\linewidth]{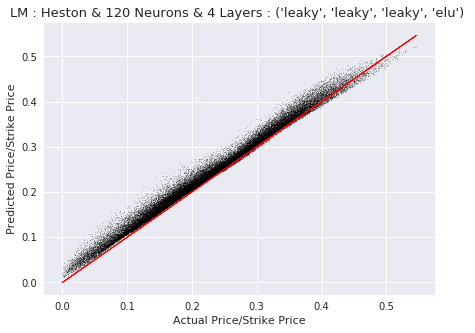}
   \end{minipage}
   \end{center}
   \caption{Performance of Pricing Machine under GBMSA Model - EC}\label{Fig:Data7}
\end{figure}

As for one example, Figure \ref{Fig:Data7} shows the performance of the architecture for the GBMSA model under in-sample test data, deep-out-of-the-money and longer maturity extrapolation data.

Furthermore, we have also studied the case where the training data set is generated with deep-out-of-the-money European call option prices under the GBM model. Therefore, by keeping all other parameter ranges the same, we generated the training data set with $S_0/K$ between 0.6 and 0.8. Table \ref{table:EC-3} provides the summary of the empirical results for the corresponding trained model under the GBM model. As in the previous tables, we check our trained model for interpolation and longer maturity cases. We also check our model for the case where $S_0/K$ is between 0.8 and 1.2, and it is under Out-Of-The-Money label in Table \ref{table:EC-3}. When the results are compared with the results in Table \ref{table:EC-1}, it is seen that In-Sample results are very similar. However, it is also seen that extrapolation results are better in Table \ref{table:EC-3}. 
Therefore, it can be concluded that although we trained the model with deep-out-of-the-money option prices, the neural network provides good predictions when the option prices are out-of-the-money. It is important to note that by out-of-the-money, we mean the moneyness between 0.8 and 1.2.

\begin{table}[!htb]
\begin{center}
\scalebox{0.85}{
\begin{tabular}{|c|c|c|c|c|}\hline
 \textbf{Training set size      }&  40000 & 80000 & 160000 & 240000\\\hline
  \multicolumn{1}{|c}{{\small \textbf{In-Sample}}}&\multicolumn{4}{|c|}{} \\ \hline
MSE & 0.000138 & 0.000042 & 0.000029& 0.000018   \\\hline
$R^2$ &  99.30\% & 99.79\% & 99.85\%& 99.91\%    \\\hline
\multicolumn{1}{|c}{{\small \textbf{Out-Of-The-Money}}} & \multicolumn{4}{|c|}{}\\\hline
MSE & 0.000166 & 0.000034 & 0.000020& 0.000014   \\\hline
$R^2$ &  98.09\% & 99.61\% & 99.78\%& 99.84\%  \\\hline
\multicolumn{1}{|c}{{\small \textbf{Longer Maturity}}} & \multicolumn{4}{|c|}{}\\\hline
MSE & 0.001845 & 0.000311 & 0.000285& 0.000221   \\\hline
$R^2$ &  92.75\% & 98.78\% & 98.88\%& 99.13\%  \\\hline
\end{tabular}}
\end{center}
\caption{European Option Pricing under GBM - 2}
\label{table:EC-3}
\end{table}

\begin{figure}[!htb]
   \begin{minipage}{0.5\textwidth}
     \centering
     \includegraphics[width=0.9\linewidth]{GBM_llll.png}
   \end{minipage}\hfill
   \begin{minipage}{0.5\textwidth}
   \centering
     \includegraphics[width=0.9\linewidth]{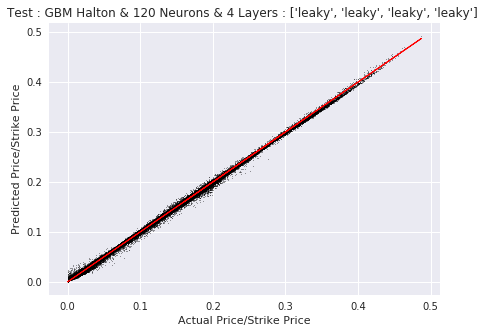}
   \end{minipage}
   \caption{Non-Quasi Approach vs. Quasi Approach under GBM Model - EC}\label{Fig:Data11}
\end{figure}

Up until now, we have only obtained results using the Non-Quasi approach. In order to see how well the Quasi approach works, we implemented it for generating data for European call options under the GBM model. Figure \ref{Fig:Data11} shows the results obtained for both Non-Quasi and Quasi approaches. It is shown that when the data is generated by using Halton sequences, the model fit is slightly better. 

\subsection{Barrier Options}
Up-and-Out Barrier put option prices are generated under 3 different models; hence, we implemented 3 different neural networks for pricing UOPs. For the GBMSA and VGSA models, we take the maximum training size as 160,000, because Monte Carlo procedure is computationally expensive and from the last section we know that the performances of the models trained with 160,000 data points and 240,000 data points are close enough. Table \ref{Table:UOBP} provides the summary of the empirical results for the trained models under the GBM, GBMSA, and VGSA models. 
\begin{table}[!htb]
\begin{center}
\scalebox{0.68}{
\begin{tabular}{|c|c|c|c|c||c|c|c||c|c|c|}\hline
& \multicolumn{4}{|c||}{{\small \textbf{GBM}}}&  \multicolumn{3}{|c||}{{\small \textbf{GBMSA}}}&  \multicolumn{3}{|c|}{{\small \textbf{VGSA}}}\\\cline{2-11}
{\small \textbf{Training set size}}&40000&80000&160000&240000&40000&80000&160000&40000&80000&160000 \\ \hline
\multicolumn{1}{|c|}{{\small \textbf{In-Sample}}} & \multicolumn{4}{|c||}{}&  \multicolumn{3}{|c||}{}&  \multicolumn{3}{|c|}{}\\\hline
MSE & 0.000159 & 0.000038 & 0.000017& 0.000016 & 0.000114 & 0.000083 & 0.000026 & 0.000149& 0.000089 & 0.000031\\\hline
$R^2$ &  95.60\% & 98.94\% & 99.54\%& 99.57\% & 96.65\% & 97.55 \% & 99.24 \% & 95.13 \%& 97.09 \% & 99.00 \% \\\hline
\multicolumn{1}{|c|}{{\small \textbf{Deep-Out-Of-The-Money}}} & \multicolumn{4}{|c||}{}&  \multicolumn{3}{|c||}{}&  \multicolumn{3}{|c|}{}\\\hline
MSE & 0.000684 & 0.000617 & 0.000766& 0.000562 & 0.001230 & 0.000965 & 0.000764 & 0.002254& 0.001733 & 0.001275\\\hline
$R^2$ &  83.64\% & 67.65\% & 91.29\%& 93.61\% & 85.43\% & 91.68 \% & 90.04 \% & 67.53 \%& 75.86 \% & 82.86 \% \\\hline
\multicolumn{1}{|c|}{{\small \textbf{Longer Maturity}}} & \multicolumn{4}{|c||}{}&  \multicolumn{3}{|c||}{}&  \multicolumn{3}{|c|}{}\\\hline
MSE & 0.000490 & 0.000107 & 0.000042& 0.000048& 0.000288 & 0.000167 & 0.000053 & 0.000163& 0.000160 & 0.000067\\\hline
$R^2$ &  86.92\% & 97.15\% &98.87\%& 98.71\%& 91.48 \% & 95.07 \% & 98.43 \% & 94.06 \%& 94.16 \% & 97.55 \% \\\hline
\end{tabular}}
\end{center}
\caption{Barrier Option Pricing under GBM, GBMSA and VGSA}
\label{Table:UOBP}
\end{table}

\begin{figure}[!htb]
   \begin{minipage}{0.5\textwidth}
     \centering
     \includegraphics[width=.9\linewidth]{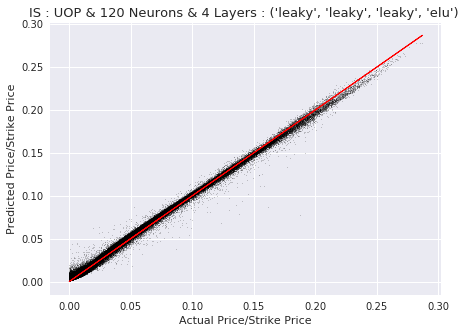}
   \end{minipage}\hfill
   \begin{minipage}{0.5\textwidth}
     \centering
     \includegraphics[width=.9\linewidth]{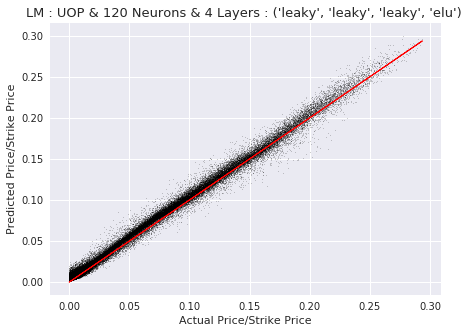}
   \end{minipage}
     \caption{Performance of Pricing Machine under GBM Model - UOBP}\label{Fig:Data8}
\end{figure}

As it is seen in Table \ref{Table:UOBP}, as the number of parameters in the model decreases, the performance of the trained model increases. As an example, Figure \ref{Fig:Data8} shows the performance of GBM model under in-sample test data and longer maturity extrapolation data.

\begin{figure}[!htb]
   \begin{minipage}{0.5\textwidth}
     \centering
     \includegraphics[width=0.9\linewidth]{UOP-llle.png}
   \end{minipage}\hfill
   \begin{minipage}{0.5\textwidth}
   \centering
     \includegraphics[width=0.9\linewidth]{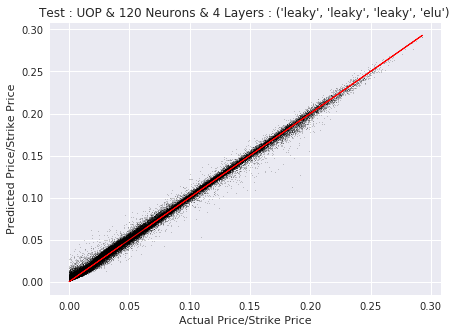}
   \end{minipage}
   \caption{Non-Quasi Approach vs. Quasi Approach under GBM Model - UOBP}\label{Fig:Data12}
\end{figure}

As in the case of European options, we also applied the Quasi approach for Up-and-Out Barrier put options under the GBM model. Figure \ref{Fig:Data12} shows the results obtained with both Non-Quasi and Quasi approaches. We can conclude that when the data is generated under the Quasi approach, the model fit is better.

\subsection{American Options}
\begin{figure}[!htb]
   \begin{minipage}{0.5\textwidth}
     \centering
     \includegraphics[width=.9\linewidth]{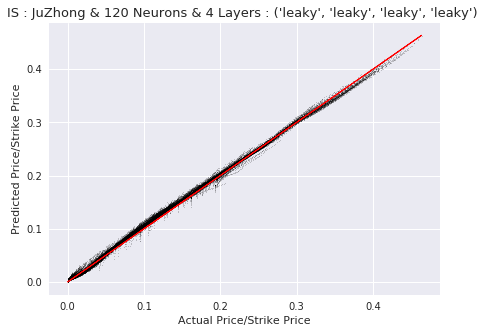}
   \end{minipage}\hfill
   \begin{minipage}{0.5\textwidth}
     \centering
     \includegraphics[width=.9\linewidth]{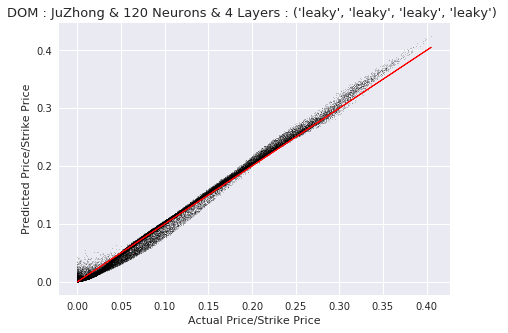}
   \end{minipage}
     \caption{Performance of Pricing Machine trained on the Ju-Zhong Approximation of GBM model - JZ}\label{Fig:Data9}
\end{figure}

American call option prices were generated under the GBM model using the Ju Zhong approximation. The empirical results from the neural network trained using this data is summarized in Table \ref{Table:AJZ}. Additionally, Figure \ref{Fig:Data9} shows the performance of this architecture, trained under the Ju-Zhong approximation of the GBM model, for in-sample test data and deep-out-of-the-money extrapolation data from left to right respectively.
\begin{table}[!htb]
\begin{center}
\scalebox{0.85}{
\begin{tabular}{|c|c|c|c|c|}\hline
 \textbf{Training set size      }&  40000 & 80000 & 160000 & 240000\\\hline
  \multicolumn{1}{|c|}{{\small \textbf{In-Sample}}}&\multicolumn{4}{|c|}{} \\ \hline
MSE & 0.000196 & 0.000187 & 0.000113& 0.000025\\\hline
$R^2$ &  99.04\% & 99.10\% & 99.45\%& 99.88\% \\\hline
  \multicolumn{1}{|c|}{{\small \textbf{Deep-Out-Of-The-Money}}}&\multicolumn{4}{|c|}{}  \\ \hline
MSE & 0.000221 & 0.000078 & 0.000098& 0.000092\\\hline
$R^2$ &  82.70\% & 93.88\% & 92.30\%& 92.79\% \\\hline
  \multicolumn{1}{|c|}{{\small \textbf{Longer Maturity}}}&\multicolumn{4}{|c|}{}  \\ \hline
MSE & 0.000788 & 0.000500 & 0.000364& 0.000812\\\hline
$R^2$ &  97.55\% & 98.44\% & 98.87\%& 97.48\% \\\hline
\end{tabular}}
\end{center}
\caption{American Option Pricing under GBM}
\label{Table:AJZ}
\end{table}

\section{Deep Neural Network vs. Recurrent Neural Network}
\begin{figure}[!htb]
   \begin{minipage}{0.5\textwidth}
     \centering
     \includegraphics[width=.9\linewidth]{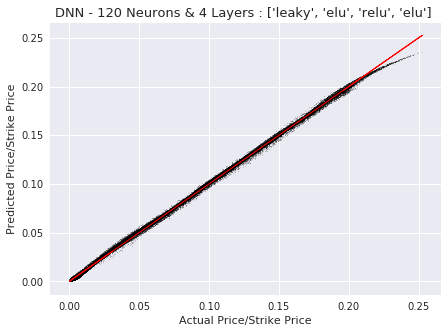}
   \end{minipage}\hfill
   \begin{minipage}{0.5\textwidth}
     \centering
     \includegraphics[width=.9\linewidth]{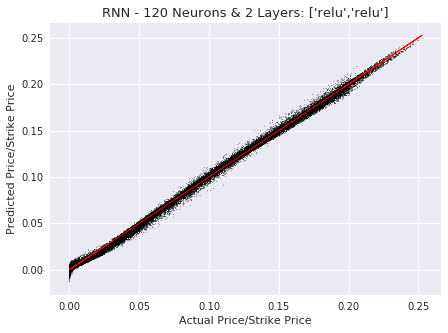}
   \end{minipage}
     \caption{DNN vs. RNN}\label{Fig:Data10}
\end{figure}

Although the trained feedforward neural network for pricing American call options performs well, as mentioned earlier it is pretty expensive to generate labels for American options. Knowing American options are strongly path-dependent options, we might benefit from recurrent neural networks in pricing them. While generating data, input parameters (except maturity) are sampled at random uniformly over the ranges defined before. For maturity, we define discrete time steps from 1 to 12 months. Figure \ref{Fig:Data10} provides a comparison between the performances of deep neural networks and recurrent neural networks. According to the numerical results, the MSE value of DNN model is slightly smaller than that of the RNN model. On the other hand, there is a big gap between training times of the two models. In our experiment training DNNs took 50\% more time than RNNs.\\

Although neural networks are proficient at storing implicit knowledge, they struggle with memorization of facts (i.e. a working memory). Human beings' ability to be able to hold memory and affiliate it with some context is critical for problem solving. Hence, a recurrent neural network, or in other words a neural network equipped with feedback loops can "not only rapidly and 'intentionally' store and retrieve facts, but also to sequentially reason with them" \cite{Goodfellow}. Enticing us to further explore the possibilities of an RNN's architecture in our future works.\\

\section{Conclusion}

In this paper, we utilized a fully connected feedforward neural network to price Vanilla, Barrier and American options under the GBM, VG, GBMSA, and VGSA regimes and explore the benefits of introducing recurrent neural networks to this task. The crux of the problem is creating labels and selecting hyper-parameters for training of the Deep Neural Network. We used sample sizes of 300,000 points for training and 60,000 points for validation. We observed with an increase in depth in layers and neurons per layer the accuracy of the predictions increases, but eventually diminishes.  Once the training is done, the model can be used for derivative pricing for market parameters both within and reasonably outside the initial range of the training set with very low error. We also compared the performances of feedforward and recurrent neural networks for American put options. We observed that training time of RNNs is almost the half of the training time of DNNs, which encourages us to explore RNNs further in our future works. All in all, once labels are created, training  and predicting options prices is very inexpensive computationally. This architecture results in speed-ups of many orders of magnitude, making it practical and easy to implement and use.

\nocite{Das}
\nocite{Komodakis}
\nocite{Spiegeleer}
\nocite{Hutchinson}
\nocite{CGMY_1}
\nocite{Weilong}
\section*{References}
\bibliography{elsarticle-template}

\begin{thebibliography}{10}
\expandafter\ifx\csname url\endcsname\relax
  \def\url#1{\texttt{#1}}\fi
\expandafter\ifx\csname urlprefix\endcsname\relax\def\urlprefix{URL }\fi
\expandafter\ifx\csname href\endcsname\relax
  \def\href#1#2{#2} \def\path#1{#1}\fi

\bibitem{FFT}
P.~Carr, D.~Madan, Option valuation using the fast fourier transform, Journal
  of computational finance 2~(4) (1999) 61--73.

\bibitem{Heston}
S.~L. Heston, A closed-form solution for options with stochastic volatility
  with applications to bond and currency options, The review of financial
  studies 6~(2) (1993) 327--343.

\bibitem{CGMY}
P.~Carr, H.~Geman, D.~B. Madan, M.~Yor, Stochastic volatility for l{\'e}vy
  processes, Mathematical finance 13~(3) (2003) 345--382.

\bibitem{American}
A.~Hirsa, D.~B. Madan, Pricing american options under variance gamma, Journal
  of Computational Finance 7~(2) (2004) 63--80.

\bibitem{Black}
M.~Scholes, et~al., The pricing of options and corporate liabilities, Journal
  of political Economy 81~(3) (1973) 637--654.

\bibitem{VG}
D.~B. Madan, P.~P. Carr, E.~C. Chang, The variance gamma process and option
  pricing, Review of Finance 2~(1) (1998) 79--105.

\bibitem{Hirsa}
A.~Hirsa, Computational methods in finance, CRC Press, 2016.

\bibitem{Goodfellow}
I.~Goodfellow, Y.~Bengio, A.~Courville, Y.~Bengio, Deep learning, Vol.~1, MIT
  press Cambridge, 2016.

\bibitem{LeCun}
Y.~LeCun, B.~E. Boser, J.~S. Denker, D.~Henderson, R.~E. Howard, W.~E. Hubbard,
  L.~D. Jackel, Handwritten digit recognition with a back-propagation network,
  in: Advances in neural information processing systems, 1990, pp. 396--404.

\bibitem{Asian}
J.~Vecer, M.~Xu, Pricing asian options in a semimartingale model, Quantitative
  Finance 4~(2) (2004) 170--175.

\bibitem{Soltanolkotabi}
M.~Soltanolkotabi, A.~Javanmard, J.~D. Lee, Theoretical insights into the
  optimization landscape of over-parameterized shallow neural networks, arXiv
  preprint arXiv:1707.04926.

\bibitem{Soudry}
D.~Soudry, Y.~Carmon, No bad local minima: Data independent training error
  guarantees for multilayer neural networks, arXiv preprint arXiv:1605.08361.

\bibitem{li2017convergence}
Y.~Li, Y.~Yuan, Convergence analysis of two-layer neural networks with relu
  activation, in: Advances in Neural Information Processing Systems, 2017, pp.
  597--607.

\bibitem{Kawaguchi}
K.~Kawaguchi, Deep learning without poor local minima, arXiv:1605.07110v3.

\bibitem{Choromanska}
A.~Choromanska, M.~Henaff, m.~Mathieu, B.~A. Gerard, Y.~LeCun, The loss
  surfaces of multilayer networks, arXiv:1412.0233v3.

\bibitem{Kawaguchi2}
K.~Kawaguchi, J.~Huang, L.~P. Kaelbling, Effect of depth and width on local
  minima in deep learning, arXiv preprint arXiv:1811.08150.

\bibitem{Allen-Zhu}
Z.~Allen-Zhu, Y.~Li, Z.~Song, A convergence theory for deep learning via
  over-parameterization, arXiv preprint arXiv:1811.03962.

\bibitem{kingma2014adam}
D.~P. Kingma, J.~Ba, Adam: A method for stochastic optimization, arXiv preprint
  arXiv:1412.6980.

\bibitem{Ruder}
S.~Ruder, An overview of gradient descent optimization algorithms, arXiv
  preprint arXiv:1609.04747.

\bibitem{fang2009pricing}
F.~Fang, C.~W. Oosterlee, Pricing early-exercise and discrete barrier options
  by fourier-cosine series expansions, Numerische Mathematik 114~(1) (2009) 27.

\bibitem{Glorot10understandingthe}
X.~Glorot, Y.~Bengio, Understanding the difficulty of training deep feedforward
  neural networks, in: Proceedings of the thirteenth international conference
  on artificial intelligence and statistics, 2010, pp. 249--256.

\bibitem{Glorot2}
X.~Glorot, A.~Bordes, Y.~Bengio, Deep sparse rectifier neural networks, in:
  Proceedings of the fourteenth international conference on artificial
  intelligence and statistics, 2011, pp. 315--323.

\bibitem{Das}
R.~Culkin, S.~R. Das, Machine learning in finance: The case of deep learning
  for option pricing, Journal of Investment Management 15~(4) (2017) 92--100.

\bibitem{Komodakis}
S.~Zagoruyko, N.~Komodakis, Wide residual networks, arXiv preprint
  arXiv:1605.07146.

\bibitem{Spiegeleer}
J.~De~Spiegeleer, D.~B. Madan, S.~Reyners, W.~Schoutens, Machine learning for
  quantitative finance: fast derivative pricing, hedging and fitting,
  Quantitative Finance 18~(10) (2018) 1635--1643.

\bibitem{Hutchinson}
J.~M. Hutchinson, A.~W. Lo, T.~Poggio, A nonparametric approach to pricing and
  hedging derivative securities via learning networks, The Journal of Finance
  49~(3) (1994) 851--889.

\bibitem{CGMY_1}
P.~Carr, H.~Geman, D.~B. Madan, M.~Yor, The fine structure of asset returns: An
  empirical investigation, The Journal of Business 75~(2) (2002) 305--332.

\bibitem{Weilong}
W.~Fu, A.~Hirsa, A fast method for pricing american options under the variance
  gamma model, 2019.

\end{thebibliography}

\end{document}